\pdfoutput=1

\documentclass[final,5p,times,twocolumn]{elsarticle}

\usepackage{amssymb}

\journal{Journal of Solid State Chemistry}

\begin{document}

\begin{frontmatter}



\title{Effect of Fe excess on structural, magnetic and superconducting properties of single-crystalline Fe$_{1+x}$Te$_{1-y}$Se$_{y}$}


\author[label1]{R. Viennois \fnref{1}}
\author[label1]{E.Giannini}
\author[label1]{D. van der Marel}
\author[label2]{R. \v{C}ern\'{y}}
\address[label1]{D\'epartement de Physique de la Mati\`ere Condens\' ee,
Universit\'e de Gen\`eve, 24 Quai Ernest-Ansermet, 1211 Geneva,
Switzerland}
\address[label2]{Laboratoire de Cristallographie, Universit\' e de Gen\`eve, 24 Quai Ernest-Ansermet, 1211 Geneva,
Switzerland}

\fntext[1]{Present adress: Institut Gerhardt, Universit\' e Montpellier 2, place Eug\`ene Bataillon
F-34095 Montpellier, France.}

\begin{abstract}
Single crystals of Fe$_{1+x}$Te$_{1-y}$Se$_{y}$ have been grown with a controlled Fe excess and Se doping, and the crystal structure has been refined for various compositions. The systematic investigation of magnetic and superconducting properties as a function of the structural parameters shows how the material can be driven into various ground states, depending on doping and the structural modifications. Our results prove that the occupation of the additional Fe site, Fe2, enhances the spin localization. By reducing the excess Fe, the antiferromagnetic ordering is weakened, and the superconducting ground state is favored. We have found that both Fe excess and Se doping in synergy determine the properties of the material and an improved 3-dimensional phase diagram is proposed.

\end{abstract}


\end{frontmatter}


\section{Introduction}
\label{1}

The discovery of superconductivity in Fe-based pnictides, \emph{RE}OFeAs (\emph{RE} = rare earth), at temperatures as high as 55K \cite{Kamihara08,Ren08} has roused a research rush on new superconducting materials that contain Fe and share a common structural feature, i.e. layers of almost ideal PbO-like tetrahedra (see \cite{Ishida09} for a review and references therein). To date, five families of Fe-based superconductors have been found: \emph{RE}OFeAs, ("1111", \emph{RE}=rare earth) \cite{Kamihara08}, \emph{A}Fe$_{2}$As$_{2}$ ("122", \emph{A}=alkaline earth) \cite{Rotter08}, LiFeAs ("111") \cite{Wang08}, Fe(Se,\emph{Ch}) ("11", \emph{Ch}=S, Te) \cite{Hsu08,Yeh08} and the most recently discovered "21311" family of Sr$_{2}$\emph{M}O$_{3}$Fe\emph{Pn} (\emph{M}=Sc,V,Cr and \emph{Pn}=pnictogen) \cite{Ogino09}. Most of the undoped compounds of these families undergo a magnetic transition at low temperature, accompanied by a concomitant structural one, either at the same or slightly higher temperature \cite{Ishida09,Stanev08}. The stripe-like antiferromagnetic (AFM) order is largely believed to be due to the nesting of the Fermi surface that drives spin-density wave (SDW) ordering \cite{Stanev08}. Upon doping, either with electrons or holes, the conditions of nesting are progressively less well verified and superconductivity occurs.

Within the family of "11" binary iron chalcogenides, pure FeSe exhibits superconductivity below T$_{c}$=8\,K \cite{Hsu08}. The superconducting state exists over quite a wide range of Te-doping in the Fe(Se,Te) solid solution (up to 90\% Te substitution for Se in polycrystalline samples) with a maximum T$_{c}$ of $\simeq$15\,K \cite{Yeh08,Fang08}. However, pure FeTe is not superconducting and the two end compounds of the "11" solid solution, FeSe and FeTe, even if structurally isomorphic, reveal to be substantially different.
Both are off-stoichiometric, but whereas the off-stoichiometry in FeSe$_{1-y}$ is preferably ascribed to Se-deficiency \cite{Margadonna08,Pomjakushina09,McQueen09a}, in Fe$_{1+x}$Te excess Fe atoms occupy an additional site in the Fe-Te plane \cite{Fruchart75,Sales09,Liu09}.
In Fe$_{1+x}$Te both magnetic (AFM) and structural (tetragonal to monoclinic) transitions occur at the same temperature T$_{N}\simeq$ 67\,K \cite{Fruchart75}, with the propagation vector aligned at 45$^\circ$ from the nesting vector \cite{Bao09}. In the similar compound FeSe$_{1-y}$, however, only the structural transition (tetragonal to orthorhombic) occurs between 70 and 100\,K and no magnetic ordering is observed at low temperature \cite{Margadonna08,Pomjakushina09,McQueen09a,McQueen09b}. The structural transition in pure FeSe$_{1-y}$ is reported to disappear in Fe-rich compositions, in which superconductivity is not found \cite{McQueen09a}. Such a transition is therefore related to the occurrence of the superconducting state. In contrast, in superconducting Fe$_{1+x}$(Te,Se) the low-temperature lattice is still tetragonal \cite{Horigane09}. Finally, the superconducting states of the pure FeSe$_{1-y}$ and Te-doped Fe$_{1+x}$(Te,Se) are reported to present important differences \cite{Fang08}. All these evidences point to the importance of the "11" compounds for understanding the role of spin fluctuations in Fe-based superconductors, as well as the key role of excess Fe in both the structural and physical properties of Fe$_{1+x}$Te$_{1-y}$Se$_{y}$. A systematic investigation of structural, magnetic, and superconducting properties as a function of both the Se- and Fe-content is mandatory, and is the aim of the present work. Single crystals of Fe$_{1+x}$Te$_{1-y}$Se$_{y}$ have been recently grown by several authors \cite{Sales09,Liu09,Chen09,Wen09,Khasanov09} and the role of excess Fe in increasing the charge localization has been recognized \cite{Liu09}. However, little attention has been paid to the actual Fe and Se doping levels in a more realistic 3-D phase diagram, and how different ground states can be driven by Fe and Se doping is not yet understood. We report the first systematic study on crystal growth, structural, and magnetic properties of Fe$_{1+x}$Te$_{1-y}$Se$_{y}$ crystals, with a controlled Fe and Se composition. We clearly demonstrate how the crystal structure is affected by doping, and how such structural changes are related to superconductivity and magnetism in "11" iron-chalcogenides.

\section{Experimental}
\label{2}

 Two series of crystals of Fe$_{1+x}$(Te,Se) have been grown starting from two different Fe:(Te,Se) ratios: 1:1, and 0.9:1. According to the assessed Fe-Te phase diagram \cite{Okamoto90}, the tetragonal ($\beta$-phase) Fe$_{1+x}$Te is stable for 0.04$<$x$<$0.08 and does not melt congruently. The nominal precursor compositions actually correspond to two different compositions of the Te-rich flux. For each Fe content, a series of Fe$_{1+x}$Te$_{1-y}$Se$_{y}$ samples with y ranging from y=0 to y=0.45 was prepared. The crystals were grown using the Bridgman-Stockbarger method: the precursor mixture of Fe and (Te,Se) pieces was put in an evacuated quartz tube and sealed under vacuum. The quartz tube was placed vertically in the furnace, along a vertical temperature gradient, heated at 930-960$^\circ$C (depending on the Se-content), then slowly cooled to RT at variable cooling rates from 5 to 1 $^\circ$C/hour. The largest and most homogeneous crystals were obtained at the lowest cooling rate. Crystals were easily cleaved from the as-grown boule, with the cleavage plane perpendicular to the c-axis. The use of Fe and Te pieces instead of powders enhances the purity of the precursors and prevents the formation of unwanted oxides, often observed in these materials \cite{Janaki09}. Pure polycrystalline FeSe$_{1-y}$ (nominal composition FeSe$_{0.95}$) was also prepared for comparison, by solid state reaction of Fe and Se powders at 620$^\circ$C, followed by furnace cooling down to 420$^\circ$C where the sample was kept 2 days before cooling down to RT. In the case of FeSe$_{1-y}$, the annealing sequence was repeated several times until the sample was single-phase within the resolution of powder X-ray diffraction (XRD). The quality of the crystals was checked by XRD both in a powder diffractometer (on manually ground crystals) and in a 4-circle diffractometer, either using Co or Cu K$_{\alpha}$ radiation. The chemical composition and purity were checked by Scanning Electron Microscopy (SEM) coupled to Energy Dispersive X-ray spectroscopy (EDX). Within the uncertainty of the analysis technique, the ratio Se:Te was always found to agree with the results of XRD data (see below and Table 4). Little but non negligible local fluctuations of the Se:Te ratio were observed in some samples. Only the samples with no or very low composition gradient ($\Delta$(Se at. fraction) $\leq$0.04) were used for this study. Single-phase single-crystalline samples were obtained up to y=0.45. Single-crystal XRD was performed at room temperature on selected samples with two nominal Fe contents (1 and 0.9) and three nominal Se contents (y=0, y=0.2 and y=0.3), using a Stoe IPDS II diffractometer, with Mo K$_{\alpha}$ radiation and a graphite monochromator. Structure refinement was carried out by the least-squares method based on $\mid$F$\mid^{2}$ values using the SHELX-L program \cite{SHELXS}. Details about the refinement are summarized in Tables 1, 2, and 3. Magnetic properties were investigated using a Quantum Design MPMSII Squid Magnetometer in a magnetic field of 0.2 mT and 1 T, for the superconducting and the normal state, respectively.

\begin{table*}[h!]
  \centering
  \begin{tabular}{|c||c|c|c|c|c|}
    \hline
 Nominal Composition&FeTe&FeTe$_{0.7}$Se$_{0.3}$&FeTe$_{0.8}$Se$_{0.2}$&Fe$_{0.9}$Te$_{0.7}$Se$_{0.3}$&Fe$_{0.9}$Te$_{0.8}$Se$_{0.2}$\\
    \hline
    \hline
 $y$ [refined atomic fraction of Se]&0&0.27(4)&0.21(4)&0.32(3)&0.22(4)\\
    \hline
 $x$ [refined excess of Fe]&0.087(3)&0.053(9)&0.049(9)&0.013(9)&0.035(6)\\
    \hline
 Space group&\multicolumn{5}{|c|}{$P$4/$nmm$}\\
    \hline
 $a$[\AA]&3.826(1)&3.807(3)&3.815(2)&3.803(2)&3.806(3)\\
 $c$[\AA]&6.273(3)&6.153(7)&6.187(4)&6.136(3)&6.187(6)\\
 $V$[\AA$^3$]&91.81(5)&89.2(1)&90.02(9)&88.73(8)&89.6(1)\\
    \hline
 Z&\multicolumn{5}{|c|}{2}\\
    \hline
 Wavelength&\multicolumn{5}{|c|}{0.71073 (MoK )}\\
    \hline
 Crystal shape&\multicolumn{5}{|c|}{Plate}\\
    \hline
 Crystal size [mm]&0.064$\times$0.056$\times$&0.147$\times$0.086$\times$&0.076$\times$0.043$\times$&0.178$\times$0.178$\times$&0.151$\times$0.122$\times$\\
 &$\times$0.006&$\times$0.006&$\times$0.004&$\times$0.016&$\times$0.008\\
    \hline
 Absorption correction&\multicolumn{5}{|c|}{numerical from crystal shape and size \cite{Stoe99}}\\
    \hline
 Data collection&\multicolumn{5}{|c|}{Stoe IPDS II, $\omega$ oscillation}\\
    \hline
 Detector distance [mm]&\multicolumn{5}{|c|}{80}\\
    \hline
 Exposure time [min]&10&15&20&5&6\\
    \hline
 Range; increment [$^\circ$]&0-180;1&0-180;1&0-180;1.5&0-180;1&0-180;1.5\\
    \hline
 2$\theta$ interval [$^\circ$]&\multicolumn{5}{|c|}{2.86-64.80}\\
    \hline
 &-5$\leq$h$\leq$5&-5$\leq$h$\leq$5&-5$\leq$h$\leq$5&-5$\leq$h$\leq$4&-5$\leq$h$\leq$5\\
 Range in $hkl$&-5$\leq$k$\leq$5&-5$\leq$k$\leq$5&-5$\leq$k$\leq$5&-5$\leq$k$\leq$5&-5$\leq$k$\leq$5\\
 &-9$\leq$l$\leq$8&-9$\leq$l$\leq$7&-9$\leq$l$\leq$7&-8$\leq$l$\leq$7&-7$\leq$l$\leq$8\\
    \hline
 Reflections measured&1048&1025&1030&942&929\\
    \hline
 Reflections unique&114&111&112&105&106\\
    \hline
 $R_{int}$&0.028&0.107&0.057&0.039&0.049\\
    \hline
 Refinement&\multicolumn{5}{|c|}{$|F|^2$ all unique reflections}\\
    \hline
 Data/params.&114/8&111/9&112/9&105/9&106/9\\
    \hline
 Refinement software&\multicolumn{5}{|c|}{SHELXL-97 \cite{SHELXS}}\\
    \hline
 $S$&1.202&1.274&1.257&1.276&1.482\\
    \hline
 $R_{F}$&0.017&0.068&0.064&0.043&0.064\\
    \hline
 $R_{wF^2}$&0.031&0.159&0.156&0.098&0.169\\
    \hline
Residual in difference electron-density map&-0.71,0.84&-2.42,4.46&-1.49,5.90&-1.32,3.38&-2.87,5.41\\
    \hline
 \end{tabular}
  \caption{Parameters of data collection and treatment for single crystals of Fe$_{1+x}$Te$_{1-y}$Se$_{y}$.}\label{param}
\end{table*}

\begin{table*}[h!]
  \centering
  \begin{tabular}{|c|c|c|c|c|c|c|}
    \hline
 Site&Wyckoff symbol&x&y&z&U$_{eq}$&N\\
\hline
\hline
 \multicolumn{7}{|c|}{\textbf{FeTe}}\\
    \hline
 Fe1&2a&3/4&1/4&0&0.0152(3)&1\\
    \hline
 Fe2&2c&1/4&1/4&0.718(2)&= U$_{eq}$(Fe1)&0.087(3)\\
    \hline
 Te&2c&1/4&1/4&0.28141(8)&0.0163(2)&1\\
\hline
\hline
 \multicolumn{7}{|c|}{\textbf{FeTe$_{0.7}$Se$_{0.3}$}}\\
    \hline
 Fe1&2a&3/4&1/4&0&0.021(1)&1\\
    \hline
 Fe2&2c&1/4&1/4&0.70(1)&= U$_{eq}$(Fe1)&0.05(1)\\
    \hline
 Te$_{1-N}$Se$_{N}$&2c&1/4&1/4&0.2787(4)&0.025(1)&0.27(4)\\
\hline
\hline
 \multicolumn{7}{|c|}{\textbf{FeTe$_{0.8}$Se$_{0.2}$}}\\
    \hline
 Fe1&2a&3/4&1/4&0&0.019(1)&1\\
    \hline
 Fe2&2c&1/4&1/4&0.70(1)&= U$_{eq}$(Fe1)&0.049(9)\\
    \hline
 Te$_{1-N}$Se$_{N}$&2c&1/4&1/4&0.2789(4)&0.023(1)&0.21(4)\\
\hline
\hline
 \multicolumn{7}{|c|}{\textbf{Fe$_{0.9}$Te$_{0.7}$Se$_{0.3}$}}\\
    \hline
 Fe1&2a&3/4&1/4&0&0.0179(9)&1\\
    \hline
 Fe2&2c&1/4&1/4&0.70(3)&= U$_{eq}$(Fe1)&0.013(9)\\
    \hline
 Te$_{1-N}$Se$_{N}$&2c&1/4&1/4&0.2767(3)&0.021(1)&0.32(3)\\
\hline
\hline
 \multicolumn{7}{|c|}{\textbf{Fe$_{0.9}$Te$_{0.8}$Se$_{0.2}$}}\\
    \hline
 Fe1&2a&3/4&1/4&0&0.020(1)&1\\
    \hline
 Fe2&2c&1/4&1/4&0.70(2)&= U$_{eq}$(Fe1)&0.03(1)\\
    \hline
 Te$_{1-N}$Se$_{N}$&2c&1/4&1/4&0.2779(4)&0.023(1)&0.22(4)\\
    \hline

 \end{tabular}
  \caption{Atomic coordinates, equivalent displacement parameters (\AA) and site occupancies in Fe$_{1+x}$Te$_{1-y}$Se$_{y}$.}\label{param}
\end{table*}

\begin{table*}[h!]
  \centering
  \begin{tabular}{|c|c|c|c|c|c|c|}
    \hline
 Site&U$_{11}$&U$_{22}$&U$_{33}$&U$_{12}$&U$_{13}$&U$_{23}$\\
\hline
\hline
 \multicolumn{7}{|c|}{\textbf{FeTe}}\\
    \hline
 Te&0.0161(2)&= U$_ {11}$&0.0167(3)&0&0&0\\
\hline
\hline
 \multicolumn{7}{|c|}{\textbf{FeTe$_{0.7}$Se$_{0.3}$}}\\
    \hline
 Te$_{1-N}$Se$_{N}$&0.017(1)&= U$_{11}$&0.038(1)&0&0&0\\
\hline
\hline
 \multicolumn{7}{|c|}{\textbf{FeTe$_{0.8}$Se$_{0.2}$}}\\
    \hline
 Te$_{1-N}$Se$_{N}$&0.016(1)&= U$_{11}$&0.036(1)&0&0&0\\
\hline
\hline
 \multicolumn{7}{|c|}{\textbf{Fe$_{0.9}$Te$_{0.7}$Se$_{0.3}$}}\\
    \hline
 Te$_{1-N}$Se$_{N}$&0.0166(8)&= U$_{11}$&0.031(1)&0&0&0\\
\hline
\hline
 \multicolumn{7}{|c|}{\textbf{Fe$_{0.9}$Te$_{0.8}$Se$_{0.2}$}}\\
    \hline
 Te$_{1-N}$Se$_{N}$&0.017(1)&= U$_{11}$&0.035(1)&0&0&0\\
    \hline

 \end{tabular}
  \caption{Atomic anisotropic displacement parameters (\AA$^2$) in Fe$_{1+x}$Te$_{1-y}$Se$_{y}$.}\label{param}
\end{table*}

\section{Results}
\label{3}

\subsection{Structure refinement}
\label{3.1}

\begin{figure}[ttt]
\includegraphics[width=\columnwidth]{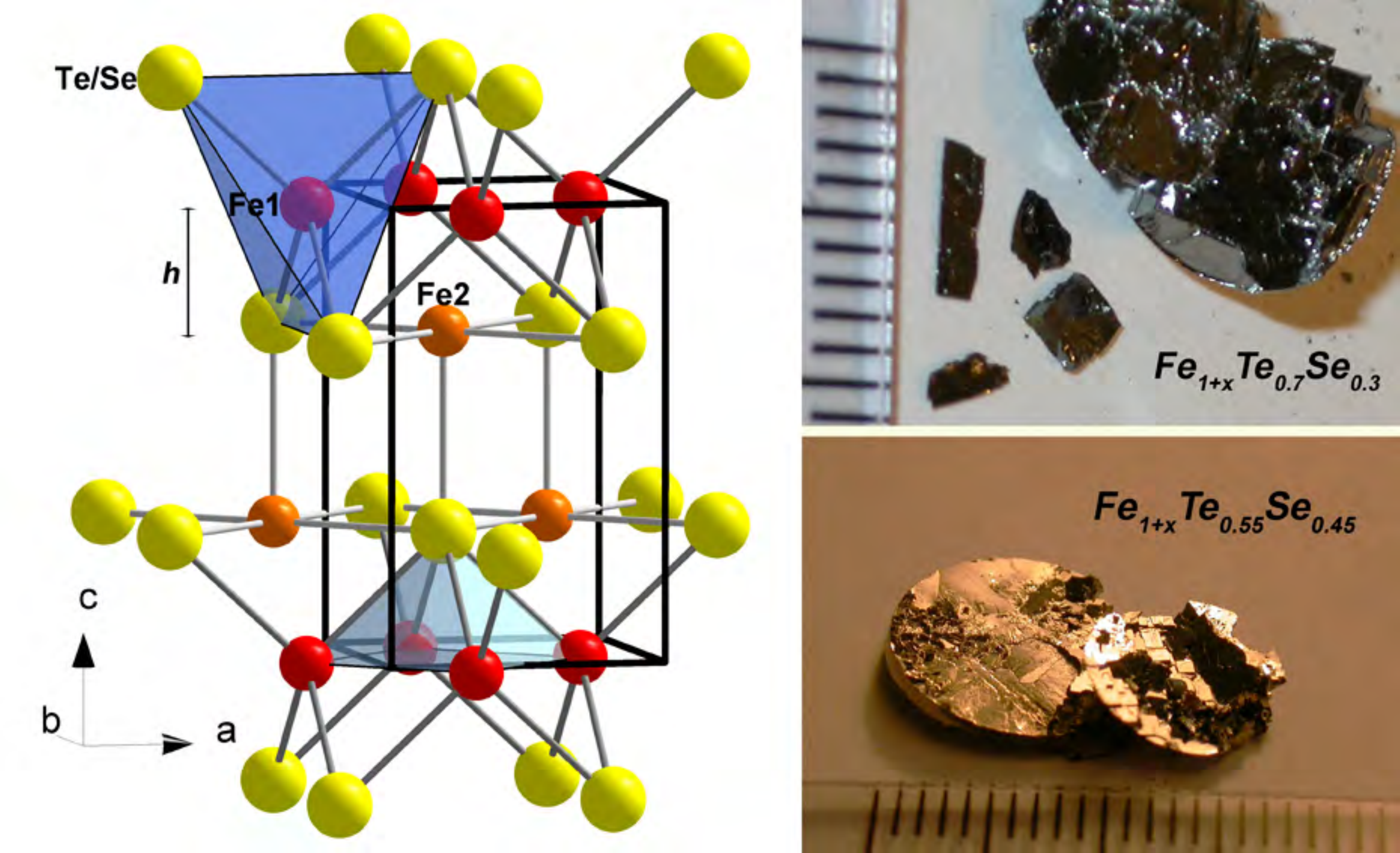}
\caption{Crystal structure of $\beta$-Fe$_{1+x}$Te, showing the tetrahedral coordination (blue) of Fe1 by four Te atoms, distance h of Fe1 from the next Te layer, and the square pyramid (light blue) TeFe$_{4}$. The pictures in the right panel show typical crystals used for this study. \label{fig:ViennoisJSSC_Fig1}}
\end{figure}

For all samples used for the single-crystal XRD study, the reflections were indexed to a tetragonal cell in the space group \emph{P}4\emph{/nmm}. The structure model was confirmed to be of the Cu$_{2}$Sb structure type as proposed by Fruchart et al. \cite{Fruchart75}, according to which the Fe atoms can occupy an additional site, in the Te-plane. (see Fig.\,\ref{fig:ViennoisJSSC_Fig1}. The figure also shows some of the crystals we measured). The occupation of the additional site turns out to be dependent on the initial composition, and to get close to zero in crystals grown from a Fe-deficient and Se-rich starting composition. The XRD data acquisition and structure refinement parameters are summarized in Table 1. The first three lines of the Table 1 compare the actual (refined) Fe and Se atomic contents and the nominal composition of various samples. The atomic positions and occupations are reported in Table 2, whereas the anisotropic displacement parameters are listed in Table 3. The samples used for this study, together with the corresponding transition temperatures, are summarized in Table 4.

\begin{table*}[h!]
  \centering
  \begin{tabular}{|c|c|c|c|c|c|}
  \hline
    \multicolumn{2}{|c|}{Composition}&Te:Se ratio from EDX&T$_{c}$,\,K&T$_{N}$,\,K&remarks\\
    Nominal&Refined&&&&\\
  \hline\hline
 FeTe&Fe$_{1.087}$Te&1:0&--&67&refined structure\\
 FeTe$_{0.9}$Se$_{0.1}$&--&0.90(1):0.10(3)&--&46&\\
 FeTe$_{0.8}$Se$_{0.2}$&Fe$_{1.049}$Te$_{0.79}$Se$_{0.21}$&0.79(2):0.21(4)&--&14&refined structure\\
 FeTe$_{0.7}$Se$_{0.3}$&Fe$_{1.053}$Te$_{0.73}$Se$_{0.27}$&0.68(2):0.32(4)&--&5&refined structure\\
 FeTe$_{0.55}$Se$_{0.45}$&--&0.55(2):0.45(4)&--&4&\\
 FeTe$_{0.50}$Se$_{0.50}$&--&--&10&--&not single phase\\
  \hline
 Fe$_{0.9}$Te$_{0.9}$Se$_{0.1}$&--&0.90(1):0.10(3)&--&22&\\
 Fe$_{0.9}$Te$_{0.8}$Se$_{0.2}$&Fe$_{1.035}$Te$_{0.78}$Se$_{0.22}$&0.76(2):0.24(4)&6&--&refined structure\\
 Fe$_{0.9}$Te$_{0.7}$Se$_{0.3}$&Fe$_{1.013}$Te$_{0.68}$Se$_{0.32}$&0.66(2):0.34(4)&11&--&refined structure\\
 Fe$_{0.9}$Te$_{0.55}$Se$_{0.45}$&--&--&14&--&not single phase\\
    \hline
  \end{tabular}
  \caption{Nominal, refined, and measured (EDX) compositions of the samples used for this study. Columns 4-5 report the corresponding superconducting and magnetic transition temperatures, as plotted in Fig.\,\ref{fig:ViennoisJSSC_Fig7}. The samples Fe$_{0.9}$Te$_{0.55}$Se$_{0.45}$ and FeTe$_{0.5}$Se$_{0.5}$, are not single-phase and were not used for structural studies, but are added for completeness.}\label{param}
\end{table*}

\begin{figure}[t]
\includegraphics[width=\columnwidth]{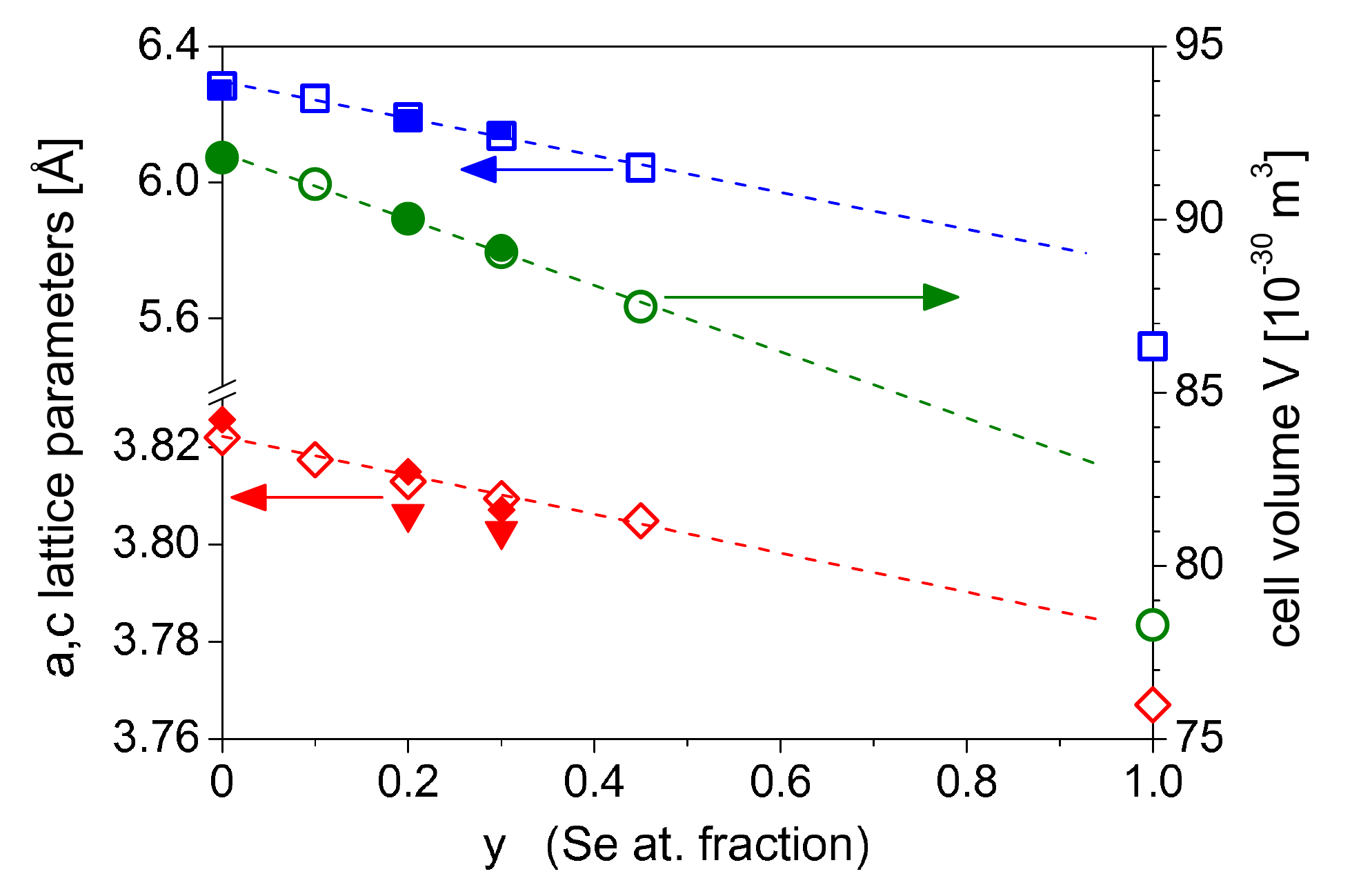}
\caption{Lattice parameters as a function of Se content. Full symbols: single crystals X-ray diffraction (triangles - low Fe excess, squares - high Fe excess). Open symbols: powder diffraction (from ground single crystals). Dashed lines show the linear dependence at low Se-doping levels, which is not obeyed by the pure FeSe. \label{fig:ViennoisJSSC_Fig2}}
\end{figure}

The lattice parameters $a$ and $c$ are reported in Fig.\,\ref{fig:ViennoisJSSC_Fig2}, plotted as a function of the Se-content, $y$. Both parameters decrease following a good linear trend upon doping in the Te-rich region. However, the lattice parameters of pure $\beta$-FeSe (added in Fig.\,\ref{fig:ViennoisJSSC_Fig2} for comparison) clearly diverge from Vegard's law, thus indicating a structural difference between $\beta$-FeTe and $\beta$-FeSe, probably related to the Se-vacancies in pure iron selenide. These results are in good agreement with those already existing in the literature \cite{Fang08,Margadonna08,Sales09,Horigane09}. In order to understand the structural and physical role of excess Fe in the additional site, we have studied the evolution of various parameters with both the refined Fe content and Se-doping. The first conclusion, resulting from the refined compositions and confirmed by EDX analysis, is that the occupancy of the second Fe site decreases when the Se substitution for Te increases. This indicates that the stability of the structure building block FeTe$_4$ is affected by both the composition parameters $x$ and $y$.

\begin{figure}[t]
\includegraphics[width=\columnwidth]{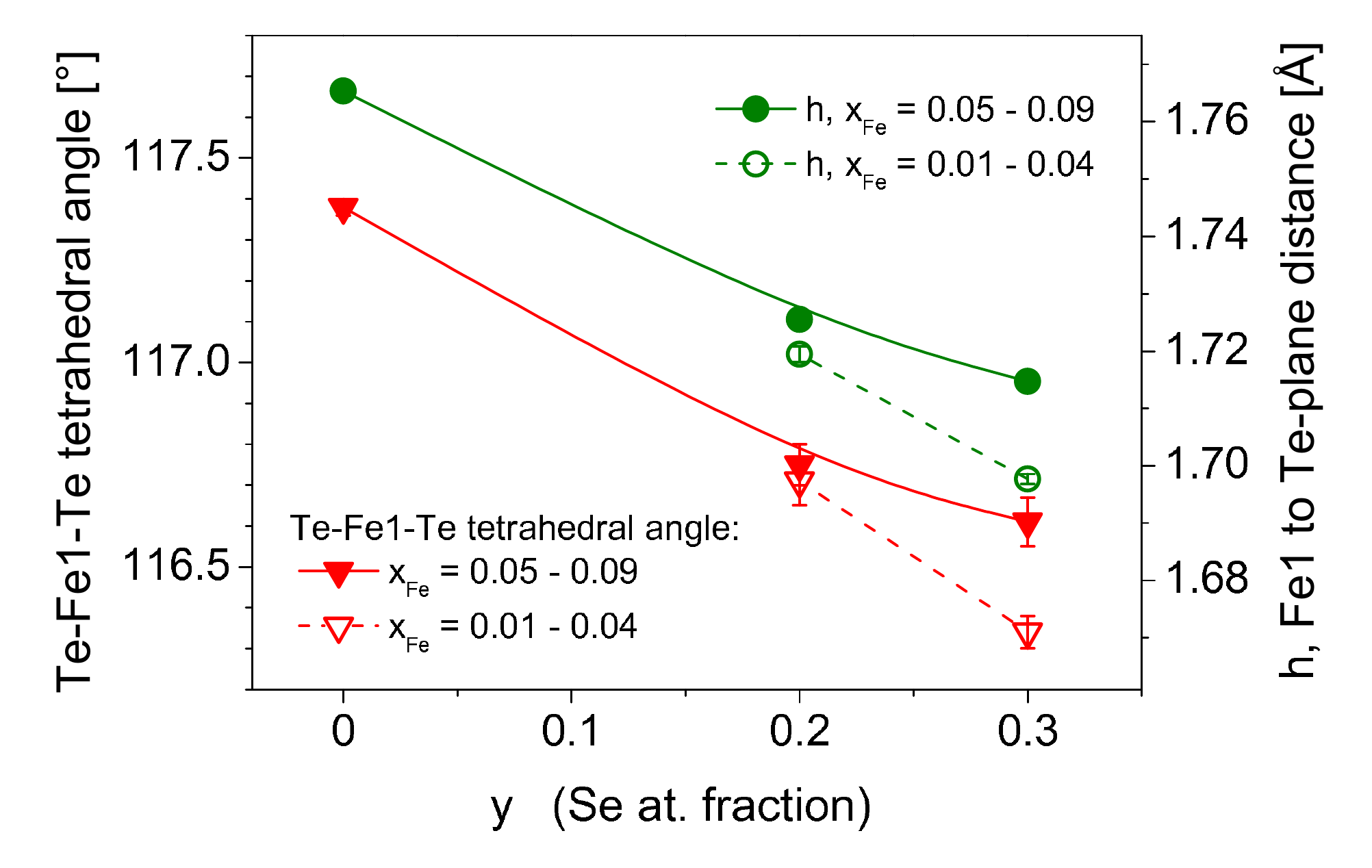}
\caption{Tetrahedral angle Te-Fe1-Te ($\triangledown$), and distance $h$ between the Fe1 atom and the Te plane ($\circ$), as a function of Se doping and for two different Fe nominal contents. Full symbols - nominal Fe=1, open symbols - nominal Fe=0.9. Legends indicate the corresponding refined excess Fe composition in the additional site. The vertical interatomic distance shortens with the Se doping, and the angle tends to the ideal one. The Fe excess influences the shrinking of the tetrahedron due to Se substitutions. \label{fig:ViennoisJSSC_Fig3}}
\end{figure}

\begin{figure}[h!]
\includegraphics[width=\columnwidth]{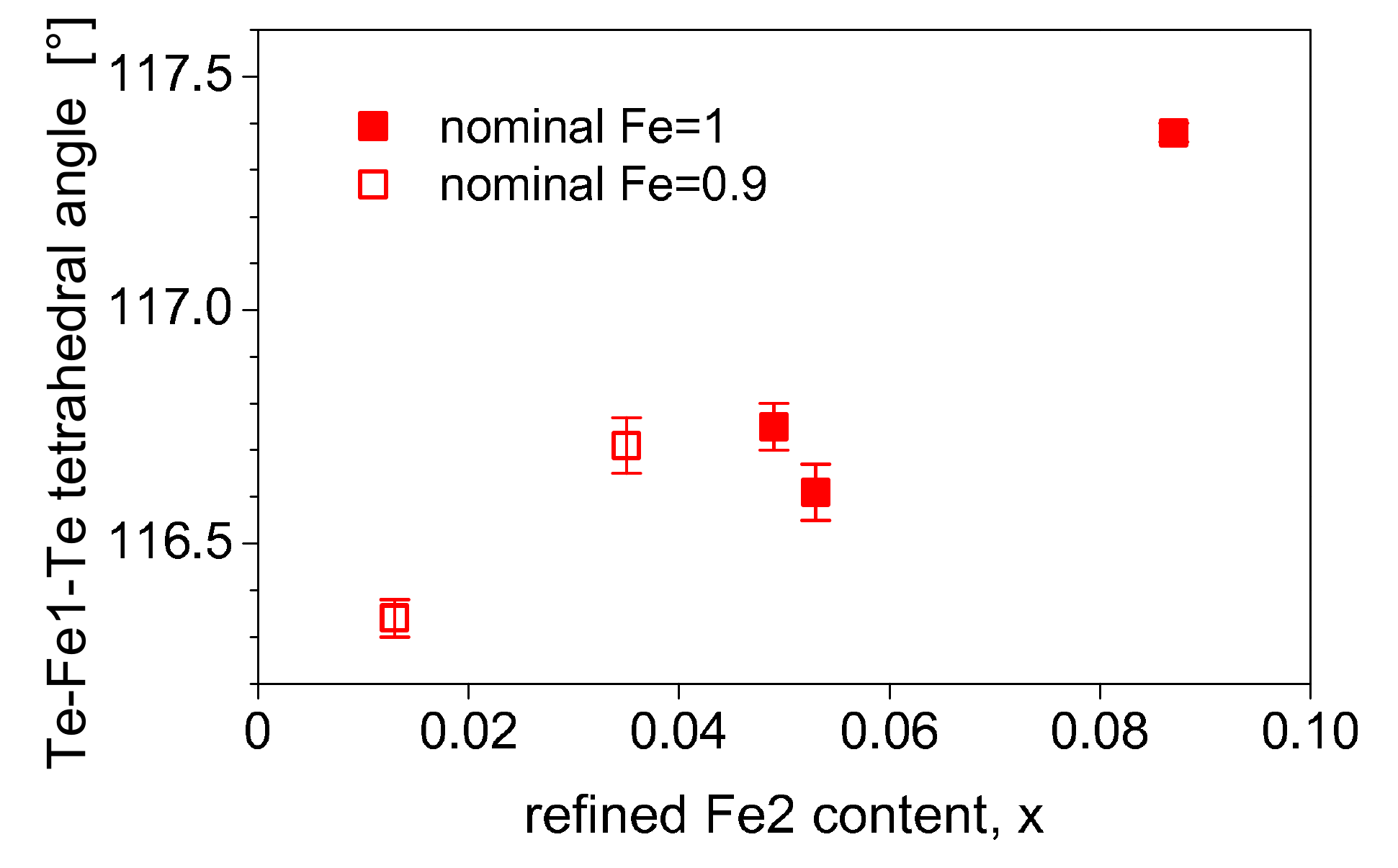}
\caption{Te-Fe1-Te tetrahedral angle as a function of the Fe2 content, $x$, as obtained from structure refinement. Full symbols - nominal Fe=1, open symbols - nominal Fe=0.9. This angle closes down with decreasing the Fe excess, regardless of the Se content. \label{fig:ViennoisJSSC_Fig4}}
\end{figure}

The corresponding structural deformation can influence the physical properties of this material. In Fig.\,\ref{fig:ViennoisJSSC_Fig3} we show the dependence of the tetrahedral angle Te-Fe1-Te (see the crystal structure and bonds shown in Fig.\,\ref{fig:ViennoisJSSC_Fig1}). This angle, $\simeq$117.4 in undoped Fe$_{1+x}$Te, is far from that of an ideal tetrahedron, 109.4$^\circ$, and gets slightly closer to it upon Se-doping. It is worth noticing that the effect of Se-doping on the tetrahedral angle is stronger when the actual (refined) Fe excess on the additional site is reduced below x=0.05 (as show by open symbols and dashed lines in Fig.\,\ref{fig:ViennoisJSSC_Fig3}). Different actual Fe compositions were obtained starting from the same nominal Fe content, but at different Te:Se ratios (see Table\,1 and Fig.\,\ref{fig:ViennoisJSSC_Fig3}). Both the $x$ and $y$ doping levels play a role in the structural stability and both have to be known and taken into account for studying the relationship between structure and properties in iron chalcogenides.
The vertical distance of the Fe1 atomic position from the Te plane obviously exhibits a similar dependence on doping as the tetrahedral angle, as is shown in Fig.\,\ref{fig:ViennoisJSSC_Fig3}. The Fe1-to-Te-plane distance, labeled $h$, shrinks with Se-doping, and its shrinking is more pronounced in samples with a lower excess of Fe. Correspondingly, the FeTe$_{4}$ tetrahedron slightly tends to the ideal shape upon reducing the Fe excess. This is shown in Fig.\,\ref{fig:ViennoisJSSC_Fig4}, where the tetrahedral angle Te-Fe1-Te is plotted as a function of the Fe excess $x$. We notice that a decreasing trend of this angle is found with decreasing the excess of iron, $x$.

\subsection{Magnetic measurements}
\label{3.2}

\begin{figure}[t]
\includegraphics[width=\columnwidth]{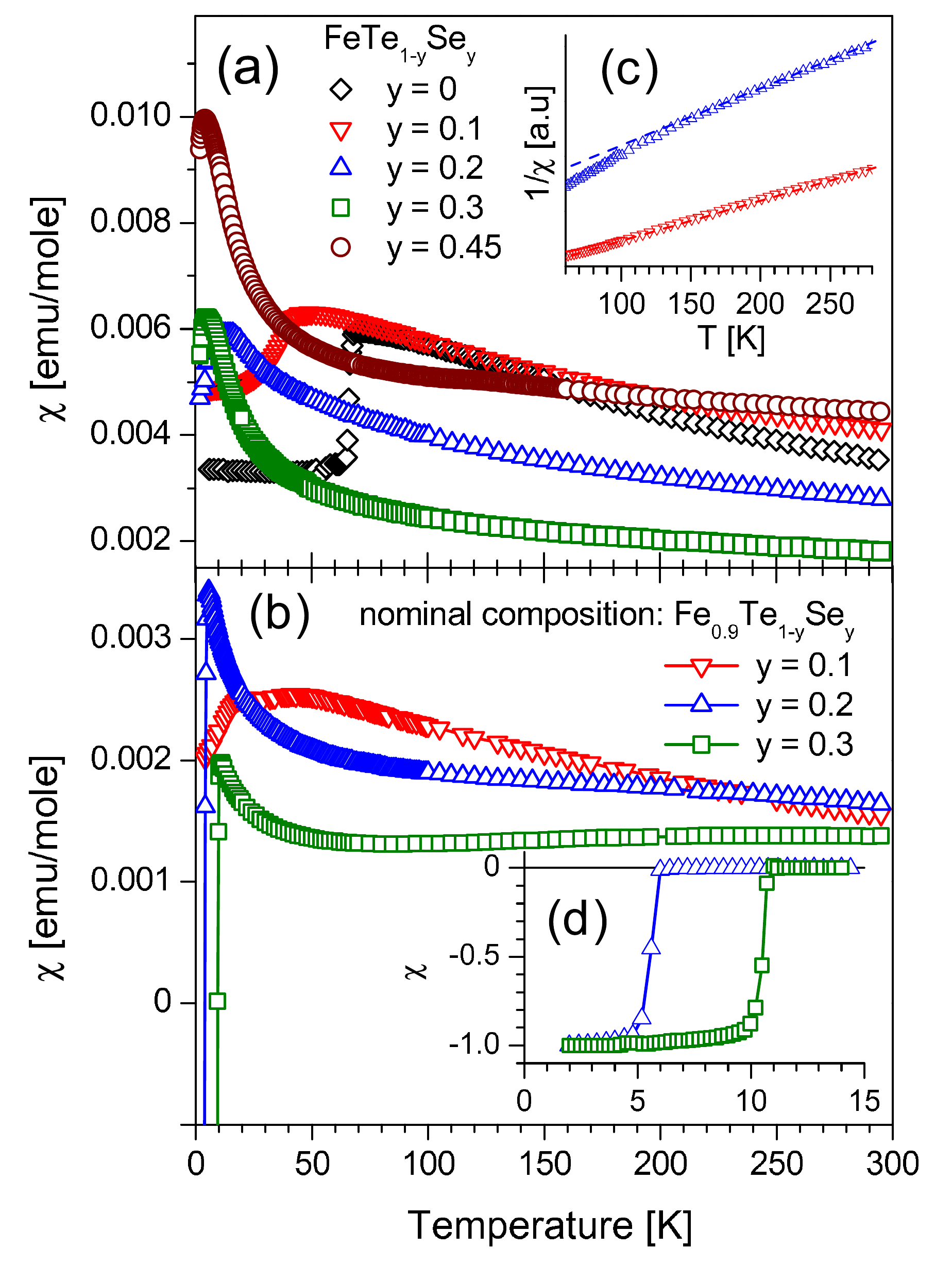}
\caption{Magnetic susceptibility $\chi(T)$ for: $(a)$ nominal composition FeTe$_{1-y}$Se$_{y}$ (Fe-rich) and $(b)$ nominal composition Fe$_{0.9}$Te$_{1-y}$Se$_{y}$ (Fe-poor), measured at $\mu$$_{0}H$=1\,T. The drop of $\chi(T)$ at the superconducting transition temperature is shown in the inset $(d)$ (at $\mu$$_{0}H$=0.2\,mT). The Curie-like decrease of $\chi$ with increasing temperature in the panel $(a)$ is less well obeyed upon increasing the Se-content, as shown in the inset $(c)$. \label{fig:ViennoisJSSC_Fig5}}
\end{figure}

The magnetic susceptibility $\chi$ is plotted as a function of temperature in Fig.\,\ref{fig:ViennoisJSSC_Fig5}, in which two series of samples corresponding to two different nominal Fe-contents, FeTe$_{1-y}$Se$_{y}$ and Fe$_{0.9}$Te$_{1-y}$Se$_{y}$, are compared ($0\leq y\leq0.45$). All FeTe$_{1-y}$Se$_{y}$ samples exhibit a Curie-Weiss-like behavior over a wide range of temperatures (see Fig.\,\ref{fig:ViennoisJSSC_Fig5}(a) and the inset 5(c)). A $\chi$ $vs.$ $1/(T-\theta)$ law is well obeyed for $y$=0 and 0.1 (Fig.\,\ref{fig:ViennoisJSSC_Fig5}(c)) . At higher Se doping levels, a more pronounced positive curvature appears below $\sim$150\,K, thus rendering the Curie-fit less reliable and the determination of the Weiss temperature $\theta$ and the effective moment $\mu_{eff}$ more uncertain. For all studied compositions, the effective moment $\mu_{eff}$ is of the order of 4\,$\mu_{B}$. For the pure Fe$_{1+x}$Te compound (actual composition Fe$_{1.087}$Te), a sharp step in the susceptibility is measured at T$_{N}$=67\, which marks the magneto-structural transition \cite{Fang08,Fruchart75,Sales09,Chen09}. When 10\% Se substitutes for Te, the high temperature behavior of the susceptibility does not change, but the transition temperatures shifts down to 46\,K. Upon increasing the Se-doping, the magneto-structural transition is suppressed (see Fig.\,\ref{fig:ViennoisJSSC_Fig5} and Table\,4). Even if filamentary or surface superconductivity is observed by transport measurements in samples with $y>0.2$ and a nominal Fe composition equal to 1, bulk superconductivity is not observed below $y=0.45$ in the samples shown in the upper panel of Fig.\,\ref{fig:ViennoisJSSC_Fig5}. A broad bulk superconducting transition is only observed at $\sim$10\,K in FeTe$_{0.5}$Se$_{0.5}$ (nominal composition), according to Ref. \cite{Sales09} (not shown here).

In Fig.\,\ref{fig:ViennoisJSSC_Fig5}(b), we report the susceptibility curves of samples with a nominal composition Fe$_{0.9}$Te$_{1-y}$Se$_{y}$, that is with a lower ($x<0.05$) Fe excess onto the Fe2 site. The magnetic susceptibility is strongly reduced at any temperature and any Se-doping in the normal state. Moreover, bulk superconductivity occurs at $y> 0.1$ as shown in Fig.\,\ref{fig:ViennoisJSSC_Fig5}(d). Superconducting transitions turned out to be as sharp as 1\,K, proving the good homogeneity of the crystals. These results clearly show that reducing the excess of Fe in Fe$_{1+x}$Te$_{1-y}$Se$_{y}$ favors the occurrence of superconductivity. A similar conclusion is also drawn by Liu et al. \cite{Liu09}, based on the comparison of two samples with different Fe-contents and a fixed Se-doping level.

\section{Discussion}
\label{4}

\begin{figure}[h!]
\includegraphics[width=\columnwidth]{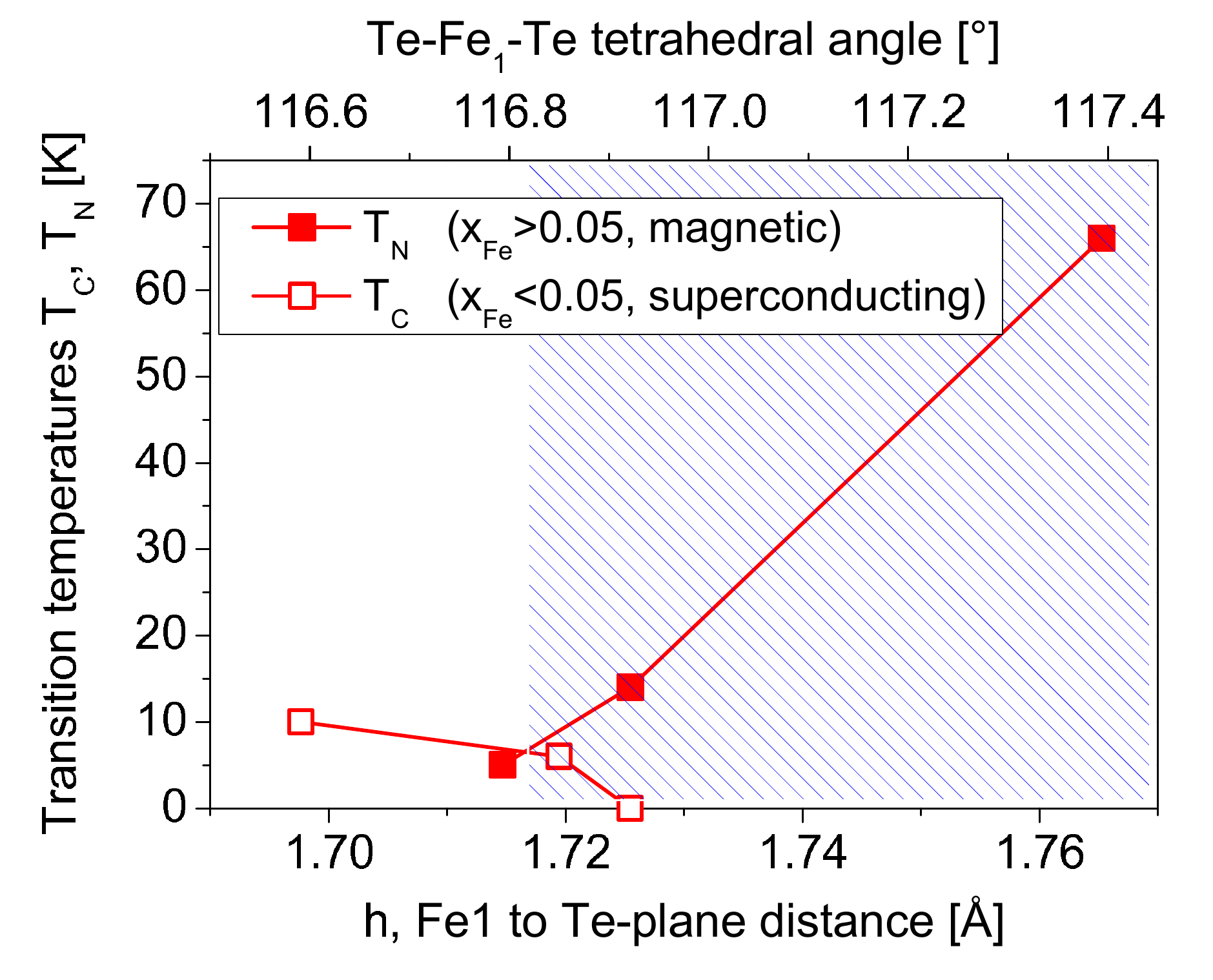}
\caption{AF and SC transition temperatures as a function of both the distance between Fe1 and Te-plane, and the tetrahedral angle. In the right hatched part, the material is antiferromagnetic (T$_{N}$ given by full symbols). In the white left site, at lower h distances and smaller angles, superconductivity occurs (T$_{C}$ given by open symbols). \label{fig:ViennoisJSSC_Fig6}}
\end{figure}

The overall decrease of susceptibility, as well as the weakening of the Curie-Weiss like contribution to $\chi(T)$ when the Fe excess is reduced, indicate that an enhancement of the localized magnetic signal is related to the Fe excess on additional sites. On the other hand, excess Fe in the Te-plane is found to be unfavorable for superconductivity. However, according to the phase diagram \cite{Okamoto90}, a little excess of Fe is needed for stabilizing the structure. Due to Se-doping for Te, less Fe is allowed to occupy the additional site, since both the effects of reducing $x$ and increasing $y$ result in shrinking and re-shaping the FeTe$_{4}$ tetrahedra. Fig.\,\ref{fig:ViennoisJSSC_Fig6} summarizes our results and points out unambiguously the two composition ranges that favor either the magnetic or the superconducting ordering. Moving from the right to the left in Fig.\,\ref{fig:ViennoisJSSC_Fig6}, the FeTe$_{4}$ tetrahedra shrink towards a lesser anisotropy and the excess Fe diminishes. Correspondingly, the magnetic (AFM) transition, marked here by T$_{N}$, the temperature of maximum in the susceptibility curves drawn in Fig.\,\ref{fig:ViennoisJSSC_Fig5}, drops to zero and the superconducting transition temperature rises. The crossover from the magnetic region (hatched in Fig.\,\ref{fig:ViennoisJSSC_Fig6}) and the superconducting one is found to happen at about $h = 1.72 \AA$.

This result is in excellent agreement with the theoretical predictions. Calculations based on Density-Functional Theory (DFT) \cite{Moon09} predict a dependence of the magnetic ordering, that is the particular kind of AFM phase, on the vertical distance between the Fe1 and Te planes. A change from a double- to single-striped AFM ordering is predicted to occur at 1.71-1.72 $\AA$, that is exactly the $h$ value at which we observed a crossover from a magnetic to a superconducting ground state. The theoretical study reported in \cite{Moon09} does not take into account any Fe excess, but only deals with Se substitutions for Te. As it results from our investigation, the distance $h$ can be tuned both by changing the Se doping and the occupation of the additional Fe2 site. In the light of our structural study and in agreement with DFT calculations, the excess Fe acts in such a way as to push the Fe1 atom away from the Te-plane and render the $(\pi,0)$ double-striped antiferromagnetism more stable. A lower excess of Fe onto the Fe2 site favors the reduction of the $h$ height, thus making the antiferromagnetism to vanish and superconductivity to arise. Johannes and Mazin demonstrated that the magnetism appears due to {\em local} Hund's rule coupling, while the particular groundstate is selected by itinerant one-electron energies \cite{Johannes09}. These energies are sensitive to the {\em h} height, and according to the model proposed by Moon et al. \cite{Moon09}, the antiferromagnetic order that is more stable at shorter $h$ would be a single-stripe $(\pi ,\pi)$ AFM order, like in Fe-based pnictides. This would imply that the mechanism that mediates pairing in both Fe-based pnictides and chalcogenides is the same. On the other hand, inelastic neutron scattering experiments on bulk superconductors FeTe$_{0.51}$Se$_{0.49}$ have revealed the same spin fluctuations in superconducting chalcogenides and pnictides \cite{Qiu09,Lumsden09}.

\begin{figure}[h!]
\includegraphics[width=\columnwidth]{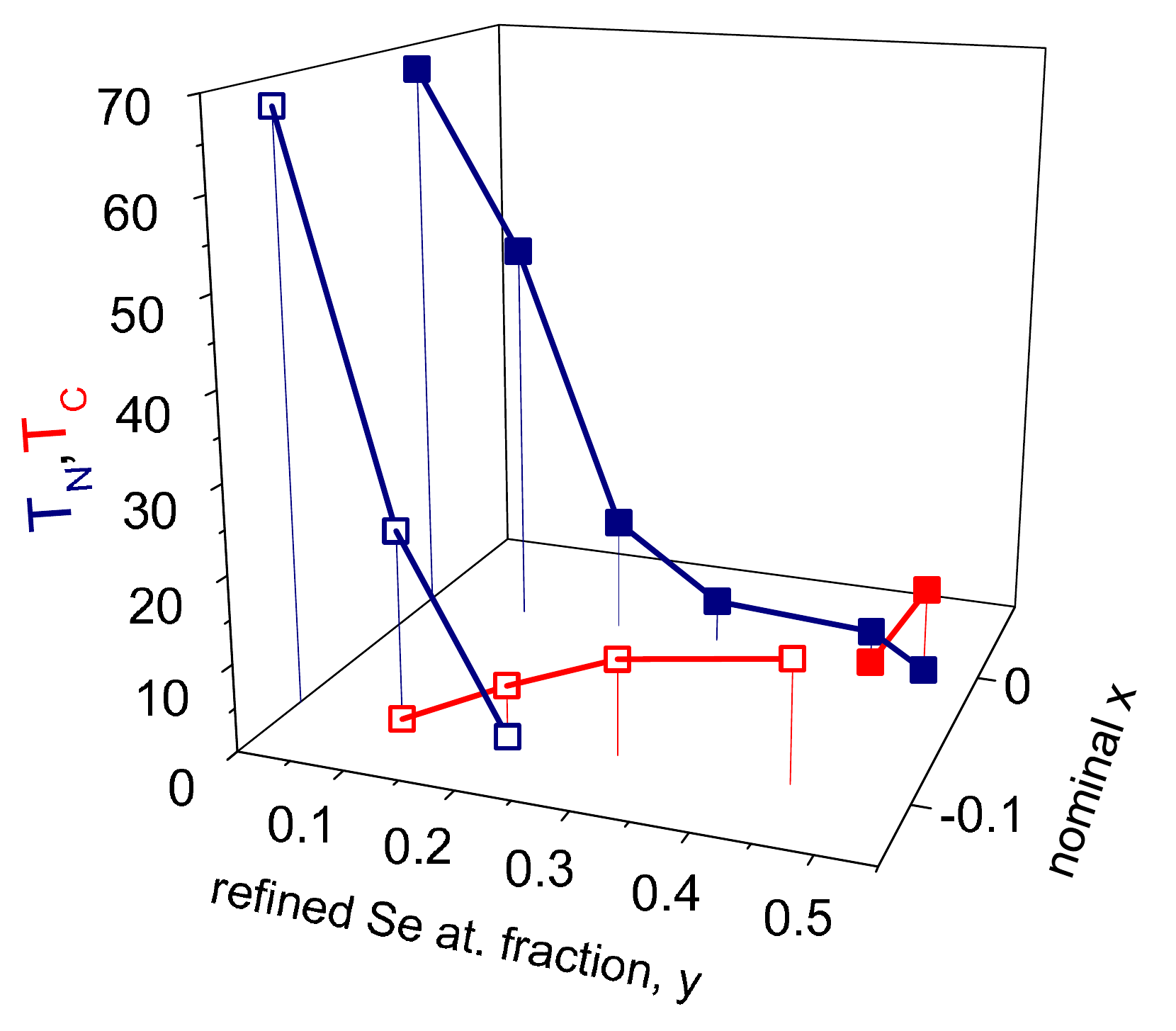}
\caption{Rough 3-D phase diagram of Fe$_{1+x}$Te$_{1-y}$Se$_{y}$. The magnetic (T$_{N}$) and superconducting (T$_{C}$) transition temperatures are represented as a function of both $x$ and $y$ doping coordinates, and listed in Table 4. \label{fig:ViennoisJSSC_Fig7}}
\end{figure}

As a general conclusion, Fe$_{1+x}$Te always forms with $x > 0$, a strongly stretched FeTe$_{4}$ tetrahedron, and exhibits double stripe antiferromagnetism with a $(\pi,0)$ propagation vector. Upon doping with Se, it is mandatory to control both the Se and Fe2 occupations in order to know when the magnetic state can switch into a superconducting one. Lowering the excess Fe concomitantly to doping with Se favors superconductivity, because it makes one particular AFM state unstable to the advantage of another. Inhomogeneities in the Fe and/or Se compositions could even make both states to coexist, and this would explain certain non-reproducibility and discrepancies existing in the literature. A more realistic view of the phase diagram should be represented in 3 dimensions, as a function of both the $x$ and $y$ doping levels. This is done in Fig.\,\ref{fig:ViennoisJSSC_Fig7} based on our experimental results, and the transition temperatures are summarized in Table 4. The source of superconductivity is not a unique doping channel, like for charge transfer in cuprates, but a multi-band effect. In this case various doping channels can contribute, and the magnetic and superconducting domes are better expressed in 3-D rather than in a conventional 2-D representation.

\section{Conclusions}
\label{5}

We have carried out a systematic study of superconductivity and magnetism in Fe$_{1+x}$Te$_{1-y}$Se$_{y}$, as a function of both the Se doping, $y$, and the excess of Fe, $x$. Single crystals have been grown of various samples and the actual compositions have been extracted from structural refinement. In such a way, any ambiguity about the actual composition, with particular regard to the excess Fe, is avoided. Our results clearly show how the occupation of the additional Fe site, Fe2, affects the magnetic and superconducting properties of the material. Evidence of enhanced spin localization is found at high Fe-content. When reducing the excess of Fe, the antiferromagnetism is suppressed and a superconducting state is favored, via a reduction of the interatomic distances and a shrinking of the FeTe$_{4}$ tetrahedron. Our experimental observations, from both structural and magnetic investigations, are in excellent agreement with the theoretical predictions from DFT calculations reported by other authors. According to the proposed model, superconducting pairing in Fe-chalcogenides and -pnictides is mediated by the same kind of spin fluctuations.

\section{Aknowledgements}

This work was partially supported by the National Center of Competence in Research "MaNEP", Materials with Novel Electronic Properties.


\bibstyle{elsart-num}
\bibliography{Viennois_JSSC_03}

\providecommand{\newblock}{}
\begin{thebibliography}{10}
\expandafter\ifx\csname url\endcsname\relax
  \def\url#1{{\tt #1}}\fi
\expandafter\ifx\csname urlprefix\endcsname\relax\def\urlprefix{URL }\fi
\providecommand{\eprint}[2][]{\url{#2}}

\bibitem{Kamihara08}
Kamihara Y, Watanabe T, Hirano M and Hosono H 2008 {\em J. Am. Chem. Soc.\/}
  {\bf 130} 3296--3297

\bibitem{Ren08}
Ren Z, Lu W, Yang J, Yi W, Shen X~L, Li Z~C, Che G~C, Dong X~L, Sun L~L, Zhou F
  and Zhao Z~X 2008 {\em Chin. Phys. Lett.\/} {\bf 25} 2215--2216

\bibitem{Ishida09}
Ishida K, Nakai Y and Hosono H 2009 {\em J. Phys. Soc. Jpn\/} {\bf 78} 062001

\bibitem{Rotter08}
Rotter M, Tegel M, Johrendt D, Schellenberg I, Hermes W and ottgen R~P 2008
  {\em Phys. Rev. B\/} {\bf 78} 020503

\bibitem{Wang08}
Wang W, Liu Q, Y-Lv, Gao W, Yang L~X, Yu R~C, Li F~Y and Jin C 2008 {\em Solid
  State Comm.\/} {\bf 148} 538--540

\bibitem{Hsu08}
Hsu F~C, Luo J~Y, Yeh K~W, Chen T~K, Huang T~W, Wu P~M, Lee Y~C, Huang Y~L, Chu
  Y~Y, Yan D~C and Wu M~K 2008 {\em Proc. Natl. Acad. Sci U.S.A.\/} {\bf 105}
  14262

\bibitem{Yeh08}
Yeh K~W, Huang T~W, Huang Y~L, Chen T~K, Hsu F~C, Wu P~M, Lee C, Chu Y~Y, Chen
  C~L, Luo J~Y, Yan Y~D~C and Wu M~K 2008 {\em Eur. Phys. Lett.\/} {\bf 84}
  37002

\bibitem{Ogino09}
Ogino H, Matsumura Y, Katsura Y, Ushiyama K, SHorii, Kishio K and Shimoyama J
  2009 {\em Supercond. Sci. Technol.\/} {\bf 22} 075008

\bibitem{Stanev08}
Stanev V, Kang J and Tesanovic Z 2008 {\em Phys. Rev. B\/} {\bf 78} 184509

\bibitem{Fang08}
Fang M, Pham H, Qian B, Liu T, Vehstedt E, Liu Y, Spinu L and Mao Z 2008 {\em
  Phys. Rev. B\/} {\bf 78} 224503

\bibitem{Margadonna08}
Margadonna S, Takabayashi Y, McDonald M, Kasperkiewicz K, Mizuguchi Y, Takano
  Y, Fitch A, Suarde E and Prassides K 2008 {\em Chem. Comm.\/} {\bf 2008}
  5607--5609

\bibitem{Pomjakushina09}
Pomjakushina E, Conder K, Pomjakushin V, Bendele M and Khasanov R 2009 {\em
  Phys. Rev. B\/} {\bf 80} 024517

\bibitem{McQueen09a}
McQueen T, Huang Q, Ksenofontov V, Felser C, Xu Q, Zandbergen H, Hor Y, Allred
  J, Williams A, Qu D, Checkelsky J, Ong N and Cava R 2009 {\em Phys. Rev. B\/}
  {\bf 79} 014522

\bibitem{Fruchart75}
Fruchart D, Convert O, Wolfers P, Madar R, Senateur J~P and Fruchart R 1975
  {\em Mat. Res. Bull.\/} {\bf 10} 169--174

\bibitem{Sales09}
Sales B, Sefat A, McGuire M, Jin R and Mandrus D 2009 {\em Phys. Rev. B\/} {\bf
  79} 094521

\bibitem{Liu09}
Liu T, Ke X, Qian B, Hu J, Fobes D, Vehstedt E~K, Pham H, Yang J, Fang M, Spinu
  L, Schiffer P, Liu Y and Mao Z 2009 {\em Phys. Rev. B\/} {\bf 80} 174509

\bibitem{Bao09}
Bao W, Qiu Y, Huang Q, Green M, Zajdel P, Fitzsimmons M, Zhernenkov M, Chang S,
  Fang M, BQian, Vehstedt E, Yang J, Pham H, Spinu L and Mao Z 2009 {\em Phys.
  Rev. Lett.\/} {\bf 102} 247001

\bibitem{McQueen09b}
McQueen T, Williams A, Stephens P, Tao J, Zhu Y, Ksenofontov V, Casper F,
  Felser C and Cava R 2009 {\em Phys. Rev. Lett.\/} {\bf 103} 057002

\bibitem{Horigane09}
Horigane K, Hiraka H and Ohoyama K 2009 {\em J. Phys. Soc. Jpn.\/} {\bf 78}
  074718

\bibitem{Chen09}
Chen G, Chen Z, Dong J, Hu W, Li G, Zhang X, Zheng P, Luo J and Wang N 2009
  {\em Phys. Rev. B\/} {\bf 79} R140509

\bibitem{Wen09}
Wen J, Xu G, Xu Z, Lin Z, Li Q, Ratcliff W, Gu G and Tranquada J 2009 {\em
  Phys. Rev. B\/} {\bf 80} 104506

\bibitem{Khasanov09}
Khasanov R, Bendele M, Amato A, Babkevich P, Boothroyd A, Cervellino A, Conder
  K, Gvasaliya S, Keller H, Klauss H~H, Luetkens H, Pomjakushina E and Roessli
  B 2009 {\em Phys. Rev. B\/} {\bf 80} 140511

\bibitem{Okamoto90}
Okamoto H and Tanner L 1990 {\em Bull. Alloy Phase Diagrams\/} {\bf 11}

\bibitem{Janaki09}
Janaki J, Kumary T, Mani A, Kalavathi S, Reddy G, Rao G and Bharathi A 2009
  {\em J. All. Com.\/} {\bf 486} 37

\bibitem{SHELXS}
Sheldrick G 1997 {\em University of Gottingen, Germany\/}

\bibitem{Moon09}
Moon C~Y and Choi H  (\textit{Preprint} \eprint{hep-ph/09092916})

\bibitem{Johannes09}
Johannes M and Mazin I 2009 {\em Phys. Rev. B\/} {\bf 79} 220510

\bibitem{Qiu09}
Qiu Y, Bao W, Zhao Y, Broholm C, Stanev V, Tesanovic Z, Gasparovic Y, Chang S,
  Hu J, Qian B, Fang M and Mao Z 2009 {\em Phys. Rev. Lett.\/} {\bf 103} 067008

\bibitem{Lumsden09}
Lumsden M, Christianson A, Goremychkin E, Nagler S, Mook H, Stone M, Abernathy
  D, Guidi T, MacDougall G, de~la Cruz C, Sefat A, McGuire M, Sales B and
  Mandrus D  (\textit{Preprint} \eprint{hep-ph/09072417})

\end{thebibliography}



\end{document}